\documentclass[a4paper,fleqn,twocolumn]{article}
\usepackage{epsfig}

\begin{document}
\author{Wojciech S\l omi\'nski\footnote{\tt email: wojteks@th.if.uj.edu.pl}
\ and
Jerzy Szwed
\\
Institute of Computer Science, Jagellonian University,
\\ 
Reymonta 4, 30-059 Krak\'ow, Poland
}
\title{\large PHENOMENOLOGY OF THE
ELECTRON STRUCTURE FUNCTION\footnote{Work supported by
the Polish State Committee for Scientific Research
(grant No. 2P03B06116).}}
\maketitle

\begin{abstract}
Advantages of introducing the electron structure function (ESF) in
electron induced processes are demonstrated. Contrary to the photon
structure function it is directly measured in such processes. At present
energies a simultaneous analysis of both the electron and the photon
structure functions gives an important test of the experimentally
applied methods. Estimates of the ESF at currently measured momenta are
given. At very high momenta contributions from $W$ and $Z$ bosons
together with $\gamma$-$Z$ interference can be observed. Predictions for
next generation of experiments are given.
\end{abstract}

\bigskip

In a series of papers \cite{ActaESF,QCDesf} we have 
presented the construction
of the electron structure function --- a useful notion in the
QCD analysis of electron induced hadron production. 
The $Q^2$ evolution equations have been constructed and 
asymptotic solutions found for the quark and gluon content
of the electron in the
leading logarithmic approximation. We included
contributions from all intermediate bosons, in particular we found 
that the $\gamma$-$Z$ interference is important
at very high energies. This  spoils the usual probabilistic
interpretation of separate $\gamma/Z/W$ structure functions.
We also found that in  certain experimental situations
the commonly
used convolution of the photon structure function and the
photon flux, used to describe the electron induced processes,
is incorrect. 

\begin{figure}[t]
\centerline{\epsfig{file=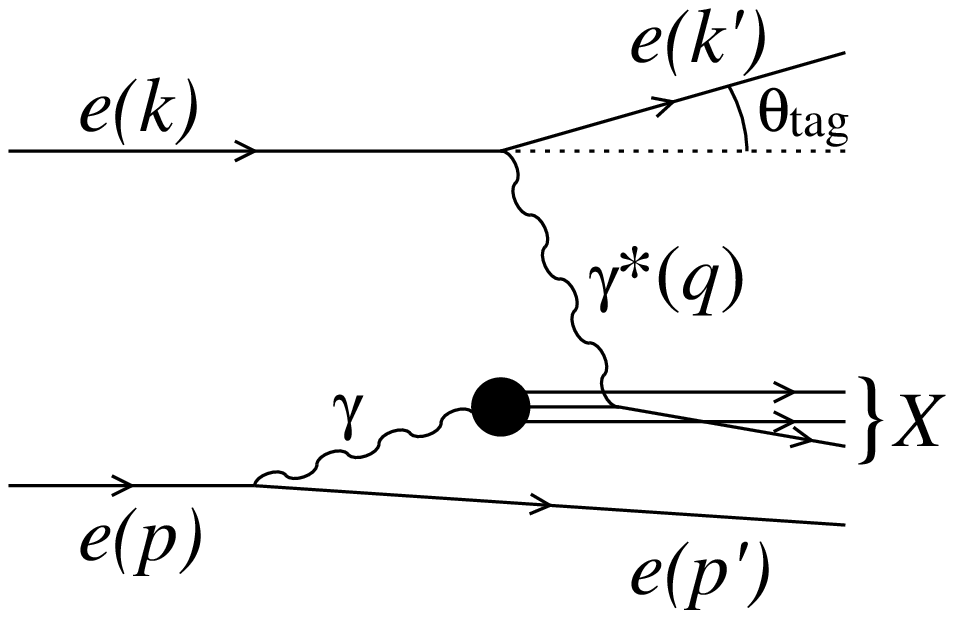,width=5cm}}
\centerline{\large (a)}
\centerline{\epsfig{file=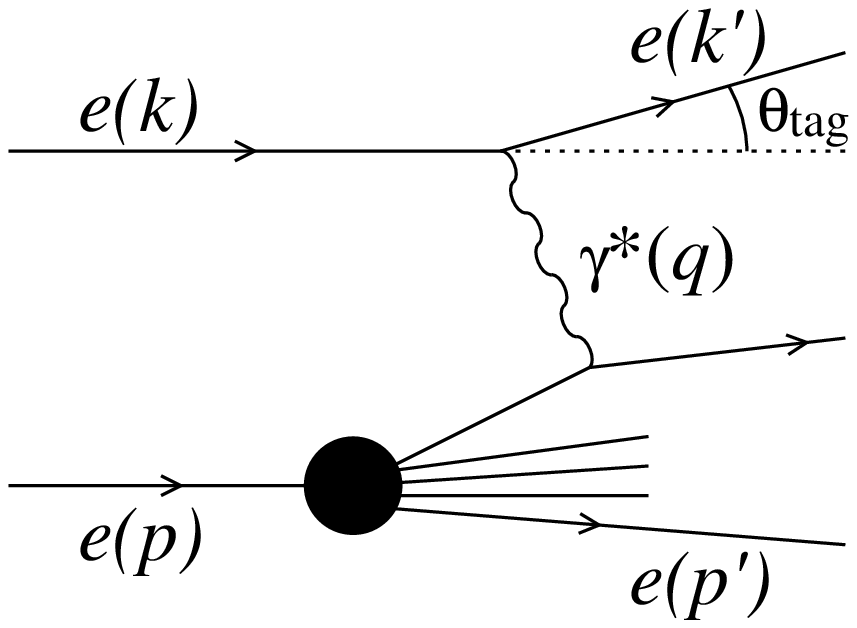,width=5cm}}
\centerline{\large (b)}
\caption{Deep inelastic scattering on a photon (a) and electron (b) target}
\label{f.DIS}
\end{figure}

In this letter we study phenomenology
of the electron structure function comparing it with the well
known approach which makes use of the photon structure function.
Theoretical framework which allows to calculate the photon structure is
known since long \cite{resph}. It appears as perturbative QCD
contribution, in addition to 
the modelled Vector Meson
Dominance  term. To measure this photonic
structure, experiments \cite{resphexp} use the electron (or positron)
beam as a source of photons. Despite precise measurement the data are
not easy to extract. 
The problem is displayed in Fig.~\ref{f.DIS}a. The
tagged (upper) electron emits a probing photon whereas the untagged
(lower) one goes nearly along the beam, emitting the target photon. 
(The situation where at very high energies
the probing boson can be also a $Z$  boson is  considered below.)
The
large scale $Q^2$ is determined by the tagged electron:

\begin{equation}
  Q^2=-(k-k')^2=2EE_{\rm tag}(1-\cos \theta_{\rm tag}),
\label{QQ}
\end{equation}
where $E$ is the initial electron energy and $E_{\rm tag}$ and $\theta_{\rm tag}$
are the energy and polar angle of the measured electron. The antitag
condition (if present) requires the virtuality of the target photon $P^2$ to be 
less than certain $P^2_{\rm max}$:
\begin{equation}
-(p-p')^2 \equiv P^2 \le P^2_{\rm max} 
\,.
\end{equation}
This photon is clearly not a beam particle and
has the energy diffused according to the equivalent photon 
spectrum:
\begin{equation}
f^e_{\gamma}(y_\gamma,P^2)=
{\alpha \over{2\pi P^2}}
\left[
  {{1+(1-y_\gamma)^2}\over{y_\gamma}}-{{2y_\gamma m_e^2}\over{P^2}}
\right]
\label{equiv}
\end{equation} 
where 
$y_\gamma$ is the photon momentum fraction,
$\alpha$ is the QED structure constant,
and $m_e$ is the electron mass.
The measured cross-section for the production of a hadronic system $X$,
expressed in terms of the 
photon structure functions $F_2^{\gamma}$ and 
$F_L^{\gamma}$ reads:
\begin{eqnarray}
\label{fot}
{{d^3\sigma _{ee\rightarrow eX}}\over dz\,dQ^2\,dx}
&=& {2\pi \alpha^2\over x^2Q^4} 
\! \int\limits_{P^2_{\rm min}}^{P^2_{\rm max}}
\!
\big[(1\!+(1-y)^2)F^{\gamma}_2(x,Q^2,P^2)
\nonumber \\
&& -y^2F_L^{\gamma}(x,Q^2,P^2)\big] 
\,f_{\gamma}^e(y_\gamma,P^2)\,dP^2
\end{eqnarray}
where 
\begin{eqnarray}
&&y = 1-(E_{\rm tag}/E)\cos ^2(\theta_{\rm tag}/2), 
\nonumber\\
&&P_{\rm min}^2 = m_e^2 y_\gamma^2/(1-y_\gamma)
\end{eqnarray}
and
$x$ ($z$)  are fractions of the parton momentum with respect
to the photon (electron).
They are related to the photon momentum fraction with respect
to the electron by
\begin{equation}
z = x y_\gamma
\end{equation}

The integral over the photon virtuality 
is usually performed by assuming $P^2 = 0$ in the 
photon structure functions 
(Weizs\"acker-Williams approximation \cite{WW}) which leads to:

\begin{eqnarray}
\label{fot2}
&&{{d^3\sigma _{ee\rightarrow eX}}\over dz\,dQ^2\,dx}
= 
{2\pi \alpha^2\over x^2Q^4} 
\nonumber\\
&&\times [(1+(1-y)^2)F^{\gamma}_2(x,Q^2,0)-y^2F_L^{\gamma}(x,Q^2,0)]
\nonumber 
\\
&&\times f_{\gamma}^{\rm WW}(z/x,P^2_{\rm max})
\end{eqnarray}
where 
\begin{eqnarray}
f_{\gamma}^{\rm WW}\! (y_\gamma,P^2_{\rm max}) &=&
{\alpha \over 2\pi}
\bigg[
{{1\! +(1-y_\gamma)^2}\over y_\gamma}\ln{{P^2_{\rm max}(1-y_\gamma)}\over{m_e^2 y_\gamma^2}}
\nonumber\\
&& -2 {1-y_\gamma \over y_\gamma}
\bigg]
\end{eqnarray}
This is how the real photon structure functions
 $F_2$, $F_L$ appear in lepton-lepton scattering.

Three remarks are important for further considerations. 
First, 
the splitting of the process into a distribution of photons 
inside electron $f_{\gamma}^{\rm WW}$ 
and that
of partons inside the photon $F_2^{\gamma}$ is an approximation. 
The optimal form of the 
equivalent photon formula is still being discussed \cite{equiv}. Even if most of 
experimental groups  choose the same 
formula, one should keep in mind
 that  the
photon structure 
function depends on this convention.
 
Second, the target photon is 
always off-shell and its virtuality is experimentally not measured. 
Although the equivalent photon distribution is peaked
 at minimum (nearly zero)
virtuality, treating the photon as 
real  is another approximation. 
One should keep in mind that the measured photon
 structure function depends on 
$x$, $Q^2$ and $P^2$, and the, usually neglected, $P^2$ 
dependence can be quite strong \cite{virt}. 
A hint that we are not measuring the real photon structure function
comes also from the analysis of the QED structure function of the photon 
where analytical solutions (and thus the $P^2$ dependence) are known \cite{Budnev}.
The data (coming from $e^+e^-\rightarrow e^+e^-\mu^+\mu^-$ scattering
  \cite{QEDsf}) agree with theory only if non-zero
 virtuality is taken into account. In the case of  QCD  the
situation is more difficult --- experiments yield the photon structure
function at virtuality $P^2$ which is not measured
and 
theory does not predict this $P^2$
dependence. 

Third,
in order to fix $x$, one is forced to measure --- in addition to the tagged 
electron --- the hadronic momenta. In fact, 
\begin{equation}
\label{xdef}
x= {Q^2 \over Q^2+P^2 + W^2}
 \approx
 {Q^2 \over Q^2 + W^2}
\,,
\end{equation}
where 
$W$ is the invariant mass of the produced hadronic system $X$. Its determination 
is more difficult than of other (tagged electron) variables since 
substantial part of hadrons is lost in the beam pipe.
The uncertainty in  the determination of the
$x$ variable is the source of large uncertainties in the analysis (unfolding procedure). 
The data are indirectly biased by theoretical assumptions. 

Many of the above problems can be avoided when we introduce the
structure function of the electron (Fig.~\ref{f.DIS}b). To see how it works 
let us first write the cross-section 
in terms of the electron structure functions $F_2^e$ and $F_L^e$:
\begin{eqnarray}
\label{ele}
{{d^2\sigma _{ee\rightarrow eX}}\over {dzdQ^2}} &=&
 {2\pi \alpha^2\over z Q^4}
\big[
(1+(1-y)^2)F^e_2(z,Q^2,P^2_{\rm max})
\nonumber \\
&& -y^2F_L^e(z,Q^2,P^2_{\rm max})
\big].
\end{eqnarray}

The structure function $F_2^e(z,Q^2,P^2_{\rm max})$ which  dominates the
cross-section at  small $y$, has simple partonic 
interpretation:
\begin{equation}
\label{F2eg}
F_2^e(z,Q^2,P^2_{\rm max})=z\sum_i e^2_{q_i} q_i(z,Q^2,P^2_{\rm max})
\,,
\end{equation}
where $e_{q_i}$ and  $q_i$ are the $i$-th quark fractional charge and density.
This is a standard deep inelastic scattering process where the
 cross-section is  related to the structure function via 
simple kinematical factors. More precisely the above defined electron
structure function corresponds exactly to the proton structure function
if no antitag condition is imposed and $P^2_{\rm max}$ goes up to
its kinematical limit of the order of $Q^2$.
(This is the `inclusive case' in our terminology \cite{ActaESF},
the electron structure function depends then on $z$ and $Q^2$ only.)

The argument $z$ --- the parton momentum fraction
with respect to the electron --- is measured, as in the standard deep inelastic 
scattering, by means of the tagged electron
variables only:
\begin{equation}
z={Q^2\over {2pq}}=
{\sin^2(\theta_{\rm tag}/2) \over E/E_{\rm tag} -\cos^2(\theta_{\rm tag}/2) }.
\end{equation}
There is no need {\it a priori} to reconstruct the hadronic mass $W$. 
In present experimental analyses one  introduces in both the photon 
and electron structure analyses 
 lower limit on $W$ because the  
reconstruction of the hadronic mass
is unreliable below this limit. 
But even so, there is an important 
difference between the case 
when $W$ (in small $W$ region) is used only as 
a selection cut to reduce background (electron structure)
and the case when in addition it is used to 
determine the $x$ variable (photon structure).  
In particular a lower cut on $W$ is much less sensitive to the unfolding
procedure than the value of $W$ itself.
The quantitative analysis of $W$ influence is discussed below.

All these features cause that
the same experiment can produce more precise 
and analysis independent
data when looking at the 
electron structure. 
What is most important
--- the electron 
structure function contains the same information about 
QCD  as the photon one
and it is known theoretically with at least the same accuracy. 
Moreover, it allows to avoid  
problems which arise in the photon structure function
at very high energies.

At present energies, where the $W$ and $Z$ bosons contributions are negligible, 
one can reanalyse the existing data in terms of the
electron structure function. This can be treated 
also as a consistency check of both photon and electron structure. 
Phenomenologically, 
having a parametrisation of the photon structure function which 
describes well the existing data,
we can predict the electron structure function 
for $Q^2 \gg P_{\rm max}^2$
by taking the 
convolution of this parametrisation with the equivalent photon spectrum:

\begin{eqnarray}
\label{konwolucja}
F^e_2(z,Q^2,P_{\rm max}^2) &=&
\int\limits_z^1 \! dy_\gamma
\!\!
\int\limits_{P^2_{\rm min}}^{P^2_{\rm max}} \!\! dP^2
f^e_{\gamma}(y_\gamma,P^2)
\nonumber\\
&&\times F_2^{\gamma}\left({z\over y_\gamma},Q^2,P^2\right)
\end{eqnarray}
 
\begin{figure}
\centerline{\epsfig{file=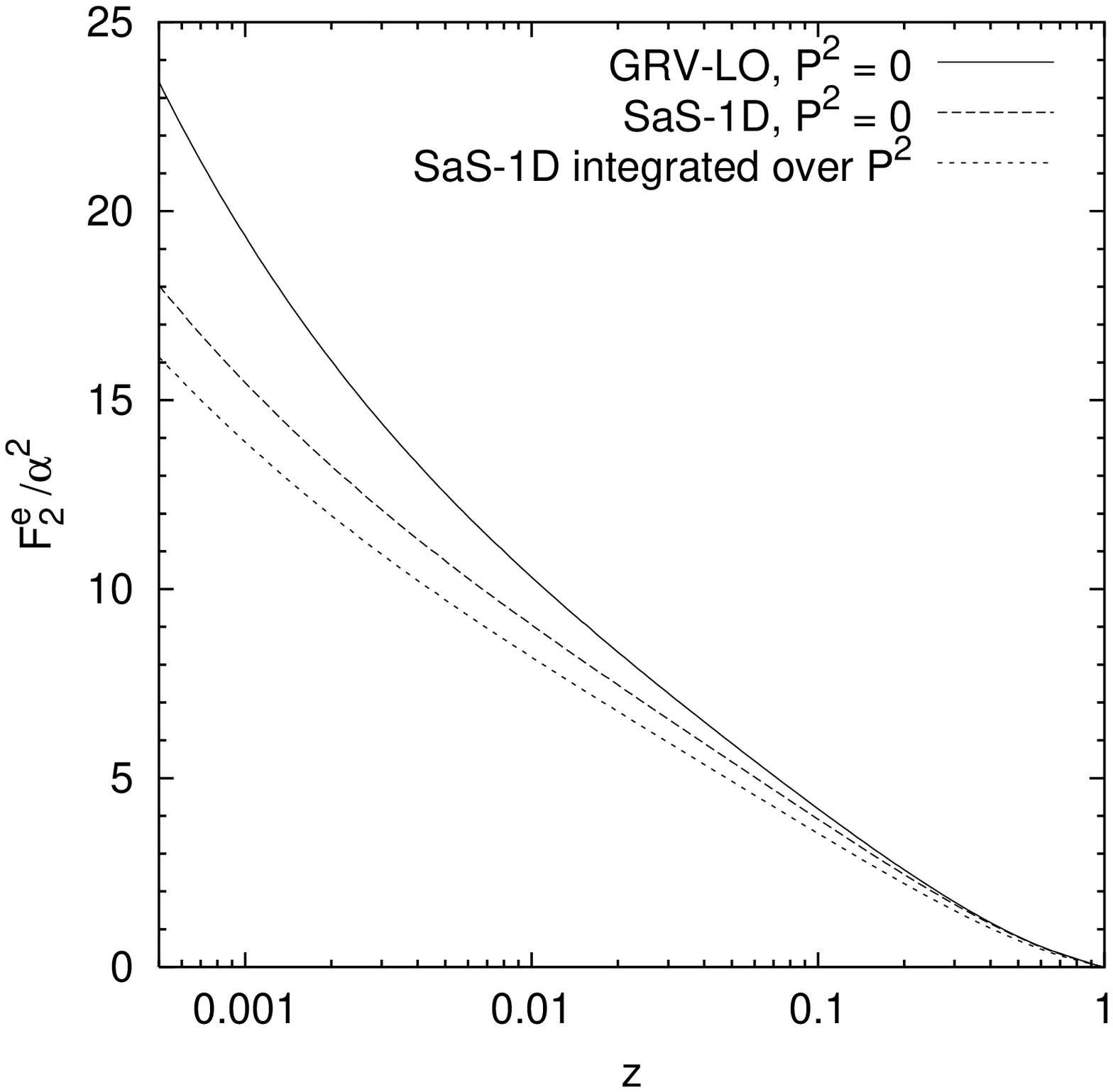,width=3in}%
\unitlength=1mm\put(-55,15){\Large a)}}
\centerline{\epsfig{file=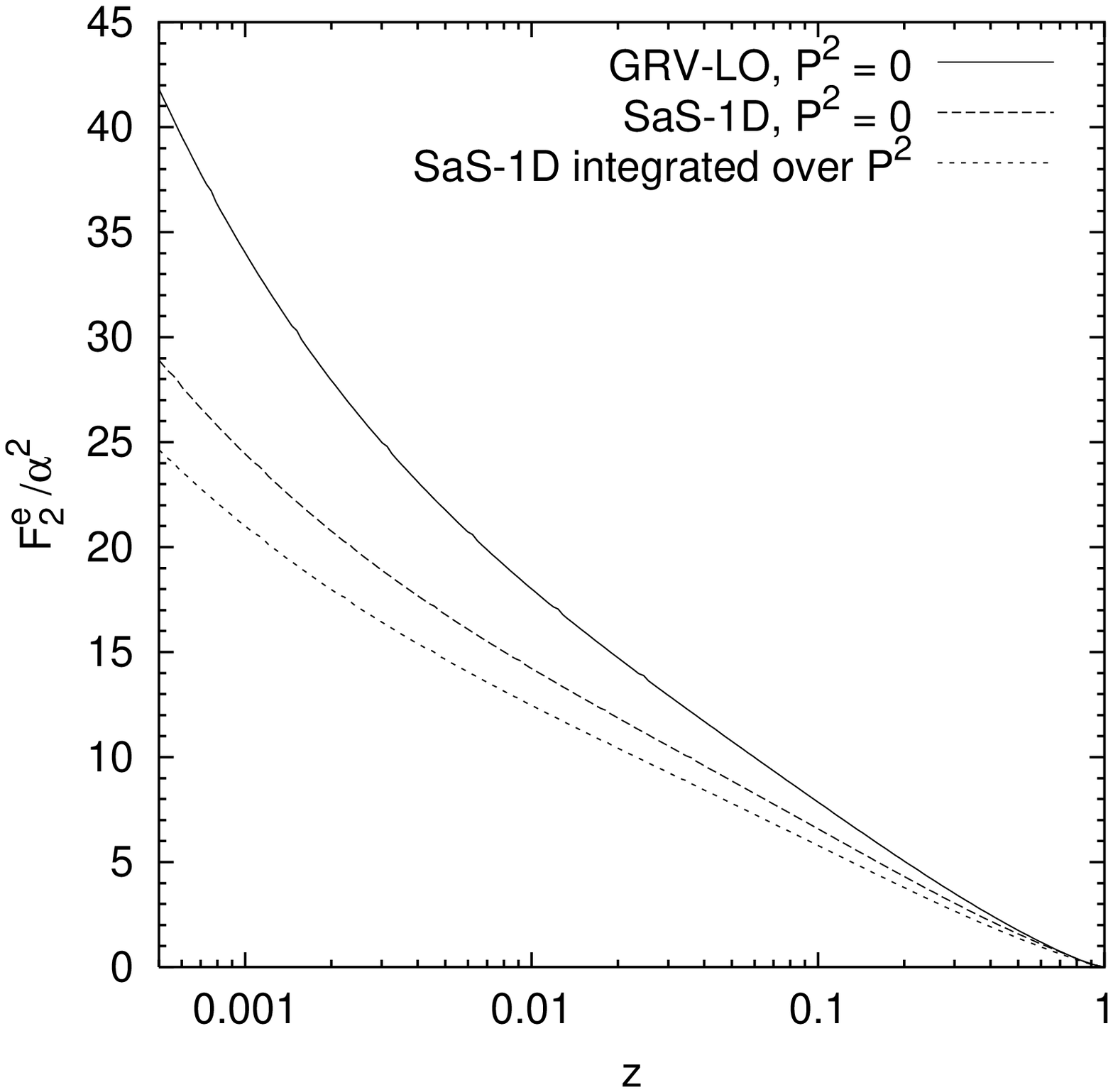,width=3in}%
\unitlength=1mm\put(-55,15){\Large b)}}
\caption{Predicted value of the electron structure function 
$F^e_2(z)/\alpha^2$ at:
a) $Q^2=17.8\;{\rm GeV}^2$, $P^2_{\rm max} = 4.5\;{\rm GeV}^2$
and 
b) $Q^2=120\;{\rm GeV}^2$, $P^2_{\rm max} = 30\;{\rm GeV}^2$
from different 
parametrisations \cite{parametrization1,parametrization2}
 of the photon structure function:
SaS-1D (broken line), GRV-LO (solid line).}
\label{f.F2e}
\end{figure}

\begin{figure}
\centerline{\epsfig{file=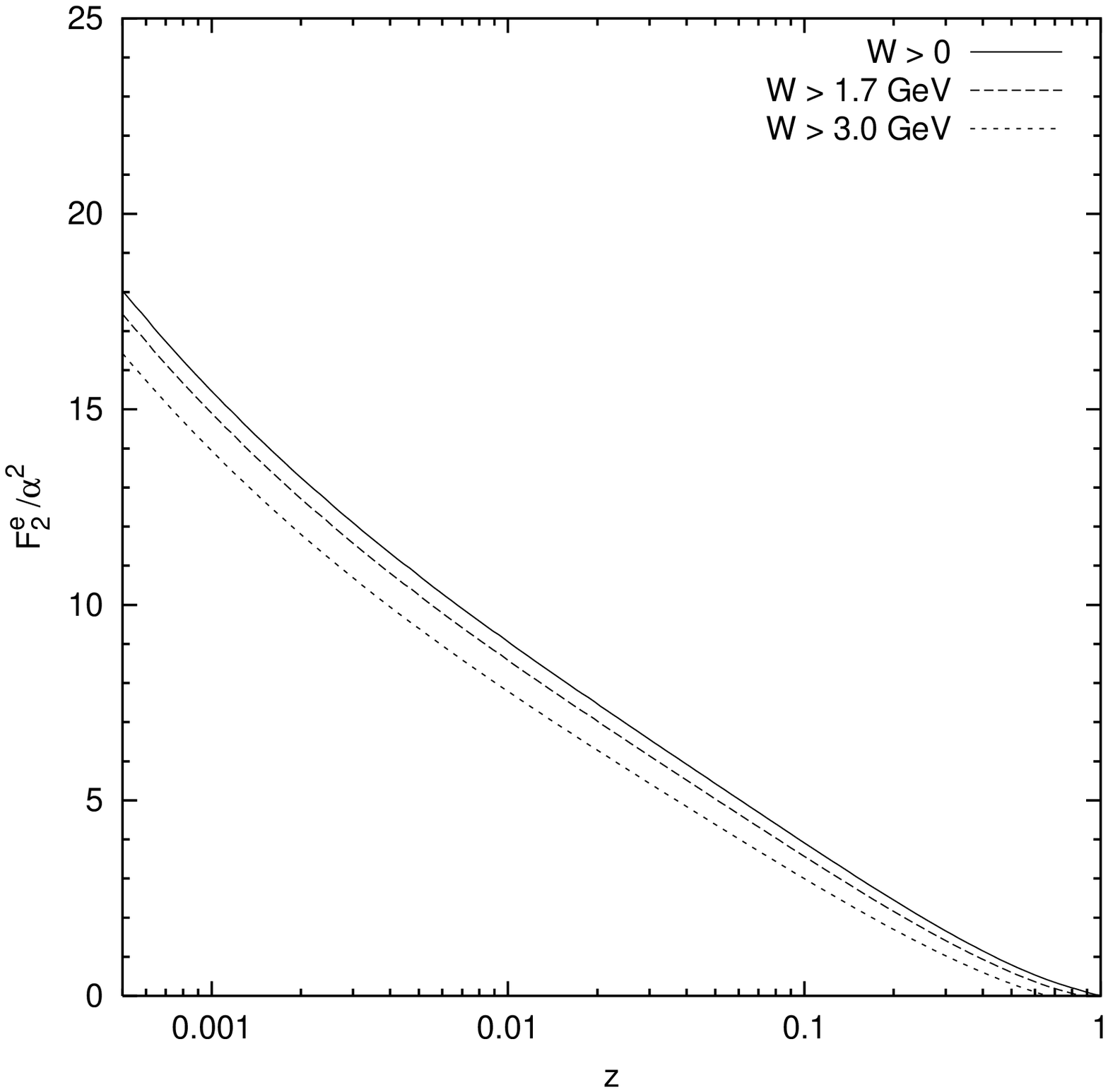,width=3in}%
\unitlength=1mm\put(-55,15){\Large a)}}
\centerline{\epsfig{file=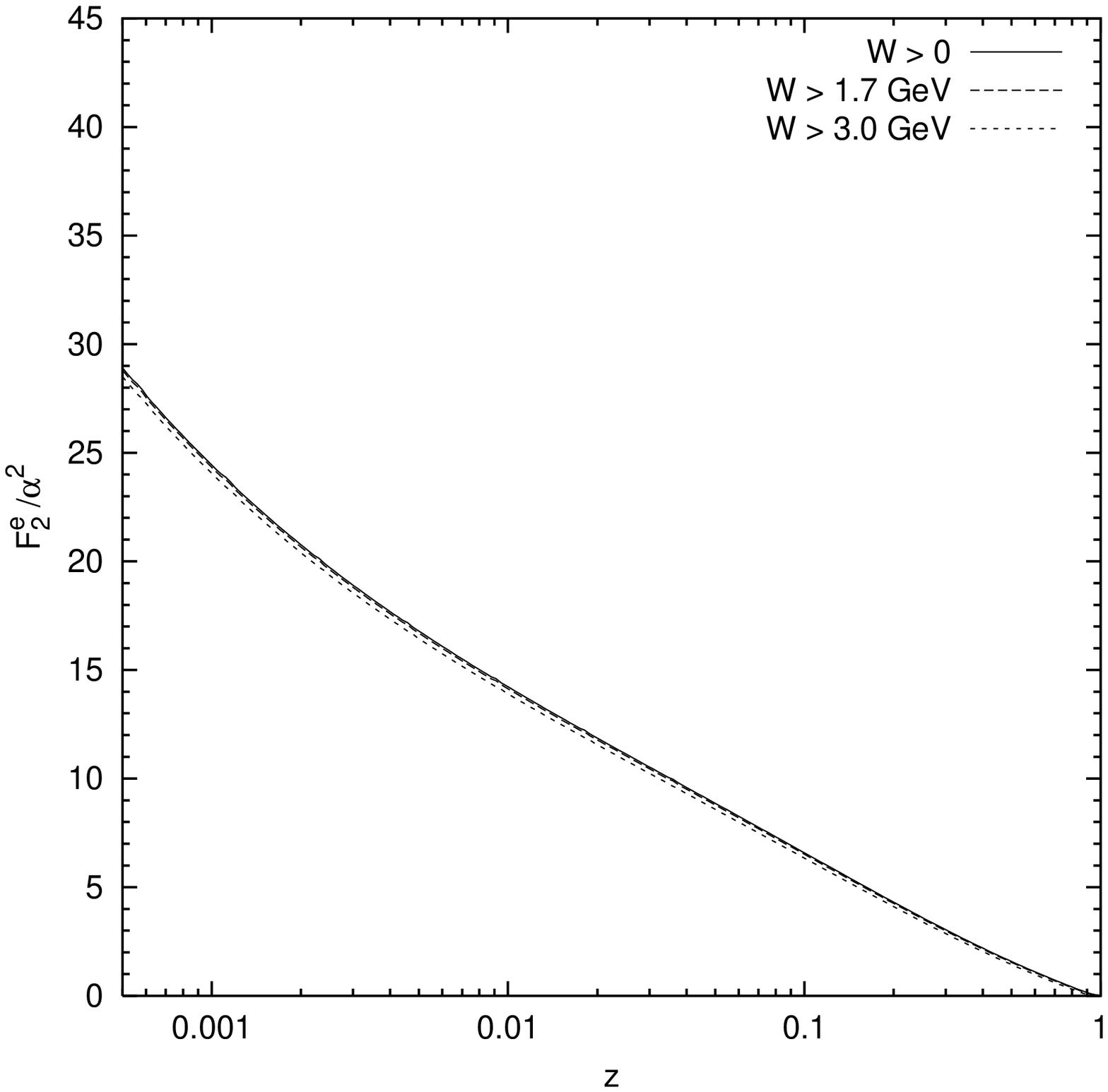,width=3in}%
\unitlength=1mm\put(-55,15){\Large b)}}
\caption{Predicted value of the electron structure function 
$F^e_2(z)/\alpha^2$ 
from SaS-1D parametrisation (solid line).
The effect on measured values when a cut on $W$ is imposed:
$W \ge 1.7$ GeV --- dashed line,
$W \ge 3.0$ GeV --- dotted line.
a) $Q^2=17.8\;{\rm GeV}^2$, $P^2_{\rm max} = 4.5\;{\rm GeV}^2$; 
b) $Q^2=120\;{\rm GeV}^2$, $P^2_{\rm max} = 30\;{\rm GeV}^2$.}
\label{fig:F2eW}
\end{figure}

Such convolution is correct when the experiments use antitag
condition (`exclusive case' in our terminology \cite{ActaESF}). 
The curves with momenta corresponding to LEP2 and TESLA/NLC/JLC experiments,
resulting from some popular parametrisations 
\cite{parametrization1,parametrization2} of real photon structure
--- {\it i.e.} $P^2 = 0$ in $F_2^{\gamma}$ of Eq.~(\ref{konwolucja})
--- are shown in Fig.~\ref{f.F2e}.
To test how significant is the (neglected above) $P^2$
dependence we also plot  the resulting curve of SaS-1D parametrisation
with  the $P^2$ dependence built in \cite{parametrization2}. 
In the latter case the integration over $P^2$ in 
Eq.(~\ref{konwolucja}) is performed numerically.
 Our previous arguments are confirmed: already at LEP2 the photon virtuality is non-negligible,
it can produce effects of the same order as differences between various parametrisations. 
The electron variables $z, Q^2$ and $P^2_{\rm max}$
are well known, the photon data do not contain one crucial information
---
the photon virtuality, at which the structure is measured.

As noticed in Ref. \cite{Forshaw}, the shape of the electron structure
function is strongly influenced by the QED part (``photon flux'').
All $F^e_2$ functions resulting from
various photon structure parametrisations 
are falling functions of $z$ and look ``similar''.
In addition the contributions at given electron momentum fraction $z=x y_\gamma$
come both from large $x$ and small $y_\gamma$ as
well as large $y_\gamma$ and small $x$.
Both these features can be
regarded as drawbacks of the electron structure function. 
The above arguments are only partly true. 
First, one should keep in mind that the variables $x$,
$y_{\gamma}$ and $z$ are not independent 
and their interplay under the integral Eq.~(\ref{konwolucja})
is specific.  To see the problem  in more detail let us 
take the Eq.~(\ref{konwolucja})
within the Weizs\"acker-Williams approximation 
($P^2_{\rm max},P^2, Q^2$ suppressed)
\begin{equation}
\label{eq:konWW}
F^e_2(z)=
\int\limits_z^1 \! dy_\gamma
f_{\gamma}^{\rm WW}(y_\gamma)
F_2^{\gamma}\left({z\over y_\gamma}\right)
\end{equation}
Noting that essentially $f_{\gamma}^{\rm WW}(y_\gamma) \propto 1/y_\gamma$
we get from Eq.~(\ref{eq:konWW}):
\begin{equation}
F_2^e(z) \propto 
\int\limits_z^1 {dy_\gamma \over y_\gamma} F_2^{\gamma}\left({z\over y_\gamma}\right)
= \int\limits_z^1 {dx \over x} F_2^{\gamma}(x)
\,.
\end{equation}
One sees that in this approximation
the $z$ dependence comes via the lower limit of integration.
Moreover the importance of the
small $x$ region under the integral (\ref{eq:konWW})
is enhanced by the $1/x$ factor.
Second, due to the same kinematics the
data points are generally shifted towards lower $z$ (as compared to $x$).
Therefore the experimental results are more accurate at small $z$, a feature
mostly welcome in the region where new effects are to be expected.
In addition, we remind  that the photon data points at low $x$ are strongest
influenced by the unfolding procedure.

As already mentioned, the present way of data analysis 
introduces a lower cut on the hadronic mass $W$.
In the photon case it causes that we do not
measure the photon structure function above certain $x$ (see Eq.~(\ref{xdef})). 
In the measurement of the
electron structure function $W$ is not needed at all (as it is not
used in the analysis of the proton structure function). 
A cut on $W$ imposed in present experiments lowers the cross-section 
in the whole $z$ range, see Fig.~\ref{fig:F2eW}.
We checked that {\it e.g.} for $Q^2 = 17.8\;{\rm GeV}^2$
the effect of the condition $W \ge 1.7$ GeV 
(with SaS-1D parametrisation in Eq.~(\ref{konwolucja}))
is less than 5\% for $z \le 0.01$, and less than 9\% for $z \le 0.1$. 
This suppression gets smaller with growing $Q^2$
({\it e.g.} at $Q^2 = 120\;{\rm GeV}^2$ it is below 1\% for $z \le 0.1$).

\begin{figure}[b]
\centerline{\epsfig{file=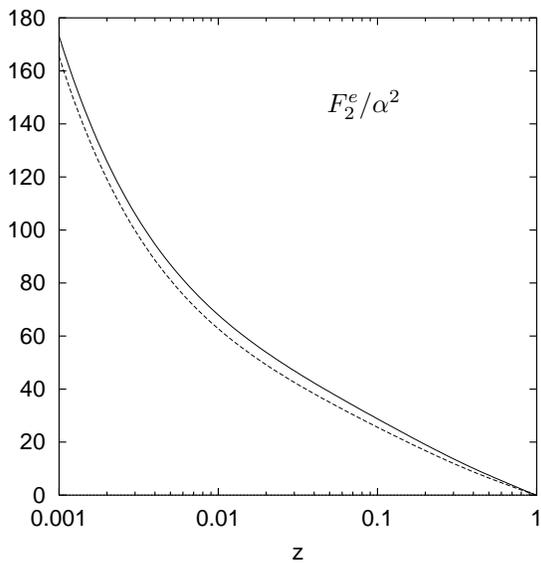,width=3in}%
\unitlength=1mm\put(-30,60){$ F_2^e/\alpha^2 $}%
}
\caption{The structure function $F^e_2(z,Q^2,P_{\rm max}^2)/\alpha^2$ at 
$Q^2=50000\;{\rm GeV}^2$ and 
$P_{\rm max}^2 = 1000\;{\rm GeV}^2$ (solid line). The contribution 
from the photon only is also shown (broken line).}
\label{fig:F2eH}
\end{figure}

The concept of the electron structure function introduces new
interesting effects at momenta much higher than
presently available. One has then to take into account not only 
the photon flux contributing to the electron structure but also those 
resulting from $Z$ 
and $W$ bosons. As shown in Ref. \cite{ActaESF} $\gamma$-$Z$ interference
comes into play and is comparable to the $Z$ contribution
itself. This means that a notion of separate gauge boson structure functions 
($\gamma, Z$ or $W$)  
looses sense and only the electron structure function preserves
probabilistic interpretation. The question is, can we observe these
effects in the next generation of experiments? 
In Fig.~\ref{fig:F2eH} we give quantitative 
estimate in the case of single tag $e^+e^-$ scattering 
at CLIC \cite{clic} momenta, choosing
$Q^2 = 10000\;{\rm GeV}^2$ and $P_{\rm max}=1000\;{\rm GeV}^2$.
In this case the picture of Fig.~\ref{f.DIS} gets modified. 
The upper (tagged) electron emits now
 both the photon and the $Z$ boson. In the calculation of the electron structure 
one has to take into account contributions from the photon, $Z$ boson, their
interference (with the antitag condition fixed by $P^2_{\rm max})$ and the $W$ boson
(no antitag condition). The presented curves are asymptotic solutions of
our electron evolution equations \cite{ActaESF}. One sees that the
effect of $Z$ and $\gamma$-$Z$ terms is of the order of $5-15\%$. We
checked that it can be enhanced to $20-25\%$ in a double tag experiment.

Let us add a few final remarks. First concerns the study of the
virtual photon structure \cite{virt} (double tag experiment). The analysis can be reformulated 
in terms of the $P_{\rm max}^2$ dependence of the electron structure function. 
Studying a real, convention independent object is  first advantage. 
Another one is the fact that
at very high virtualities the $Z$ admixture and the $\gamma$-$Z$
interference are properly taken into account.

Second is a  comment on the QED structure function of the photon.
It is obtained from the process $e^+e^- \rightarrow e^+e^-\mu^+\mu^-$
by dividing out the (approximate) equivalent photon distribution and 
assuming some  effective photon virtuality. 
The use of the QED electron structure function avoids the above approximations.
The exactly known (in given order of $\alpha$) electron 
structure function can be compared directly with the electron data.

Finally, the photon structure has been also measured 
\cite{dijets} in dijet production at HERA. Again the extraction of the $x$
variable is difficult. In addition to jets, one has to measure 
essentially  the whole hadronic system in order to obtain the photon 
energy. The data, when presented in terms of the electron 
structure, require only measurement of the two jets. 
Practically the new approach means plotting the 
dijet cross-section as a function of $z$. 
A more ambitious program would be to extract the parton densities
in the electron, as it has been done in the case of the photon
\cite{H1}
or to construct a parametrisation of parton densities inside the
electron by a direct fit to the HERA data.

To summarise, we propose to look at the electron as surrounded by a QCD
cloud of quarks and gluons (in order $\alpha^2$), very much like it is
surrounded by a QED cloud of equivalent photons (in order $\alpha$). We
argue that the use of the electron structure function in electron
induced processes has some advantages over the photon one.
Experimentally it leads to more precise, convention independent data.
Theoretically it allows for more careful treatment of all variables. It
also takes into account all electroweak gauge boson contributions,
including their interference, which will be important in the next
generation of $e^+e^-$ colliders. At present energies it should
certainly be used as a cross-check of the photon structure analysis.

The authors would like to thank Danuta Kisie\-lewska, Maria Krawczyk, Aharon Levy, 
Bogdan Muryn, Mariusz Przybycie\'n and Jacek Turnau for discussions.


\end{document}